\documentclass{PoS}

\title{GRB100814A as a member of the growing set of bursts with
sudden optical rebrightening}

\ShortTitle{GRB100814A}

\author{Massimiliano De Pasquale$^{1,2}$ N.P. Kuin$^2$, S. Oates $^2$, M. Page $^2$, S. Zane $^2$, C. Saxton$^2$,  S. Schulze$^3$,  Z. Cano$^4$,  C. Guidorzi$^5$,  P. Chandra$^6$,  D. Frail$^7$,  P. Evans$^8$,  A. Beardmore$^8$,  A. Castro-Tirado$^9$,  J. Gorosabel$^9$,  T. Fatkhullin$^{10}$,  C. Akerlof$^{11}$,  S. B. Pandey$^{12}$,  M. Im$^{13}$,  Z. L. Uhm$^1$,  Y. Urata$^{14}$, L. Huang$^{14}$, S. Pak$^{15}$, Y. Jeon$^{13}$, B. Zhang$^{1}$
\llap{$^1$}University~of~Nevada,~Las~Vegas\\
        \email{mdp@physics.unlv.edu}\\
\llap{$^2$}Mullard Space Science Laboratory, University College London Dorking, United Kingdom\\
\llap{$^3$}Faculty of Science, University of Iceland, Reykjav\'{i}k, Iceland\\
\llap{$^4$}Liverpool John Moores Liverpool, United Kingdom\\
\llap{$^5$}Deparment of Physics, Ferrara University, Ferrara, Italy\\
\llap{$^6$}Royal Military College of Canada, Canada\\
\llap{$^7$}NRAO, New Mexico, USA\\
\llap{$^8$}University of Leicester, Leicester, United Kingdom\\
\llap{$^9$}Instituto de Astrof\'{i}sica de Andaluc\'{i}a (CSIC), Spain \\
\llap{$^{10}$}Special Astrophysical Observatory, Russia\\
\llap{$^{11}$}University of Michigan, USA\\
\llap{$^{12}$}ARIES, Manora Peak, Nainital, India\\
\llap{$^{13}$}Seoul National University, Seoul, South Korea\\
\llap{$^{14}$}National Central University, Taipei, Taiwan\\
\llap{$^{15}$}Kyung Hee University Yongin, South Korea\\
}

\abstract{We present the gamma-ray, X-ray, optical and radio data for GRB100814A. At the end of the slow decline phase of the X-ray and optical afterglow, a sudden and prominent rebrightening in the optical band occurs followed by a fast decay in both bands. This optical rebrightening is accompanied by possible chromatic variations. We discuss possible interpretations, such as double component scenarios and internal dissipation mechanism, with their virtues and drawbacks. We also compare GRB100814A with other {\it Swift} bursts that show optical rebrightenings with similar properties.}

\FullConference{Gamma-Ray Bursts 2012 Conference -GRB2012,\\
		May 07-11, 2012\\
		Munich, Germany}

\begin{document}

\section{Observational data}

GRB100814A was detected by {\it Swift}/BAT (15-150 keV), {\it Konus-Wind} (20-20000 keV), {\it Suzaku}-WAM (50-5000 keV) and {\it Fermi} GBM (8-30000 keV). The 15-150 keV fluence is $9\times10^{-6}~$erg~cm$^{-2}$.  {\it Swift} XRT and UVOT promptly detected the X-ray and optical afterglow (Beardmore et al., 2010).  The redshift is z=1.44 (O'Meara et al. 2010), and the isotropic energy emitted between 10 and 1000 keV is $8\times 10^{52}$~erg.  The bright X-ray afterglow initially shows a steep decline, with the flux $F\propto t^{\alpha}$ where $\alpha=-4.65$.
After $\approx500$~s, the afterglow enters a shallow decline phase. Fitting the data from 500 s onwards with a
broken power-law model, we have $\alpha_{early,X}= -0.66 \pm 0.03$, $t_{break} = 132.7 \pm 7$~ks, $\alpha_{late,X}= -2.06\pm0.13$.
An optical source was detected by {\it Swift}/UVOT, ROTSE, Liverpool and Faulkes Telescope, Lulin Telescope,
Nordic Optical Telescope, CQUEAN at McDonald Observatory, Gran Telescopio Canarias, Calar Alto and
BTA. We present the complex photometry in Figure~1.
After an early powerlaw decay with $\alpha_0 = -0.55 \pm 0.03$, we have a conspicuous rebrightening, which
starts at $\sim15$~ks and peaks between 40 and~100 ks after the trigger, depending on frequency.
At $\sim 200$~ks, the rebrightening ends and the decay slope becomes $\alpha_3 = -1.97 \pm 0.03$.
No X-ray counterpart to the optical rebrightening is visible. Such behaviour is reminiscent of other
bursts, such as GRB081029 (Holland et al. 2012, Nardini et al. 2011), GRB080413 (Filgas et al. 2011) and GRB100621A \pos{PoS(GRB2012)104}.
The radio afterglow is detected by EVLA at 4.7 and 7.9 GHz. Modeled with a broken powerlaw, the radio
lightcurves show a peak flux of $\approx 550~\mu$ Jy at $\approx 10^6$~s and decay slope of $\alpha_{R,decay}\approx -0.8$, but different rise slopes: the 4.7 GHz flux rises with a slope $\alpha \approx 1.1$, while the 7.9 GHz has $\alpha \approx 0.2$.\\
{\it Properties of the optical rebrightening.} Fitting the individual filter lightcurves  at the optical rebrightening with a double broken powerlaw model, we find that the behaviour might be chromatic. We find possible correlations between
the peak time, the peak flux density $F(\nu_{peak})$ and peak frequency $\nu_{peak}$ (Fig.~2). The strongest
correlation is between $F(\nu_{peak})$ and $\nu_{peak}$, having a Spearman rank coefficient $-0.81$ and significance 
$\backsimeq2\sigma$. Furthermore, $\nu_{peak}$ depends on time as $\nu_{peak} \propto t^{-1.45\pm0.39}$, close to the $t^{-1.5}$ expected for the synchrotron injection frequency $\nu_m$. The powerlaw slope of the rise $\alpha_1$ is consistent with $\alpha_1\backsimeq0.6$ for all filters.
After the peak between 40 and 100 ks, we have a short plateau and then a fast flux decay. In these phases, the optical afterglow
behaves achromatically, and we fix the parameters for all lightcurves. The short plateau has a slope
$\alpha_2 \backsimeq -0.48$, and ends at a break time $t_{break,2} = 217.7\pm2.4$~ks\\
 We build 4 Spectral Energy Distributions (SEDs) of the afterglow using X-ray and optical data at 4.5, 22, 50 and 400 ks and then fit them with a simple, broken and double broken powerlaw models (see table 1). We find that the 22 and
50 ks SED fits reveal a blue spectrum, with spectral index $\beta_1 = 0.04 \div 0.3$. These values are consistent with a
synchrotron spectrum below the injection frequency $\nu_m$ or the cooling frequency $\nu_c$ (if $\nu_c < \nu_m$ ).

\section{Discussion}

The sudden and late optical rise, with no counterpart in the X-ray, can be hardly explained by a single-component external shock model. We present and describe three possible scenarios which involve two components to interpret the observed behaviour, and we outline their pros and drawbacks. 

\subsection{Double jets seen sideways}

{\it Set up}: The early X-ray and optical emission is due to a wide jet seen off-axis; the rebrightening is due to emission from a narrow jet, coaxial with the wide jet, entering the line of sight.
{\it Pros}: We find that if the observer is at angle $\theta_{obs} \approx 1.5 \theta_{wide}$ and $\theta_{narrow} \approx 0.5 \theta_{wide}$, the temporal slopes of the lightcurves can be explained. A few GRBs are thought to have double-component
ejecta (e.g. GRB030329, Berger et al. 2003; De Pasquale et al. 2009)
{\it Drawbacks}: Jets seen off-axis produce a dim afterglow. To explain the observed luminosity, the isotropic energy must be
very high ($\sim10^{56}$ erg). We also find that the density of the medium should be extremely low, of the order of a few $10^{-9}$cm$^{-3}$. To decelerate early in such a thin medium, the initial Lorentz of the outflow should be $\sim2000$. These parameters appear very improbable. The possible chromatic behaviour and the transit of a break frequency at the optical peak is not explained.
\subsection{Reverse Shock and Forward Shock interplay}
\vspace{-0.3cm}
{\it Set up}: the shallow X-ray decay may be explained by Forward Shock (FS) assuming energy injection, which powers a long-living Reverse Shock (RS; Sari \& Meszaros 2000) as well. The RS produces the early, slowly decaying optical emission.
The peak of the FS emission $\nu_m$ is initially between the X-ray and the optical band. As $\nu_m$ decreases and crosses the optical band, it produces a chromatic rebrightening. At $\approx130$~ks, a jet break occurs and both the optical and the X-ray start to decay faster. The radio peak at $\approx10^6$~s can be caused by $\nu_m$ crossing this band.
{\it Pros}: This model naturally explains the transit of a break frequency at the optical rebrightening, why this has no counterpart in the X-ray, and why optical and X-ray start to decay fast almost simultaneously.
{\it Drawbacks}: It is rather difficult to find a consistent set of physical parameters, such as the energy injection parameter and circumburst density profile, that can reproduce the early shallow decay and then the faster decay. If the break energy $E_{break}$ is indeed $\nu_m$, its time evolution between 22 and 50 ks is too rapid for the FS model.
\subsection{Internal dissipation emission}
\vspace{-0.3cm}
{\it Set up}: both the early optical and the X-ray emission are produced internally in the ejecta. When this emission
dies off, the afterglow optical peak, due to Forward Shock, is observed.
{\it Pros}: This scenario explains why the optical peak is observed towards the end of the X-ray plateau.
{\it Drawbacks}: We do not yet understand the behaviour of the internal dissipation emission, so such identification is rather {\it ad hoc}. The possible chromatic behaviour at the optical rebrightening is not clearly accounted for, nor is the late steep decay similar to that observed in the X-ray.

\begin{table*}
\begin{center}
\begin{tabular}{ccccc}
\hline \\
                          & 4.5~ks & 22~ks & 50~ks & 400~ks \\
\hline
  $\beta_1$        & $-0.54^{+0.04} _{-1.12}$ & $0.29^{+0.03} _{-0.08}$   & $0.04 ^{+0.10} _{-0.12}$  & $-0.94^{+0.02} _{-0.01}$\\
  $E_{break} (eV)$  & $845^{+600} _{-680} $  & $46.3^{+10.0} _{-3.0} $    & $3.64^{+0.40} _{-0.20}$ &  \\
  $\beta_2$        & $-1.06^{+0.12}$               & $-0.88_{-0.07}           $    & $-0.52^{+0.02} _{-0.03}$  & \\
  $E_{break,2}(eV)$& 						   &						      & $ 18.6 ^{+2.4} _{-1.6} $                   \\
  $\beta_3$        &						  &					       	&$-1.02 ^{+0.02} _{-0.03}$     \\ 
  $\chi^2$/d.o.f & $12.1/13 $                       & $55.3/34$                         & $115.6/113$         & 59.5/45 \\
\hline
\end{tabular}
\caption{Best fit parameters for the optical and X-ray SEDs of GRB100814A, with simple, broken and double broken powerlaw models. As for the double broken powerlaw model, we show the results of the fit model in which $\beta_3=\beta_2-1/2$, as predicted in the FS model if the break is the cooling frequency $\nu_c$. The spectral slope in the X-ray band is forced between -0.88 and -1.06, from the fit of the X-ray data alone.}
\label{tab_obs}
\end{center}
\end{table*}

\begin{figure*}
\begin{center}
 \includegraphics[angle=-90,scale=0.42]{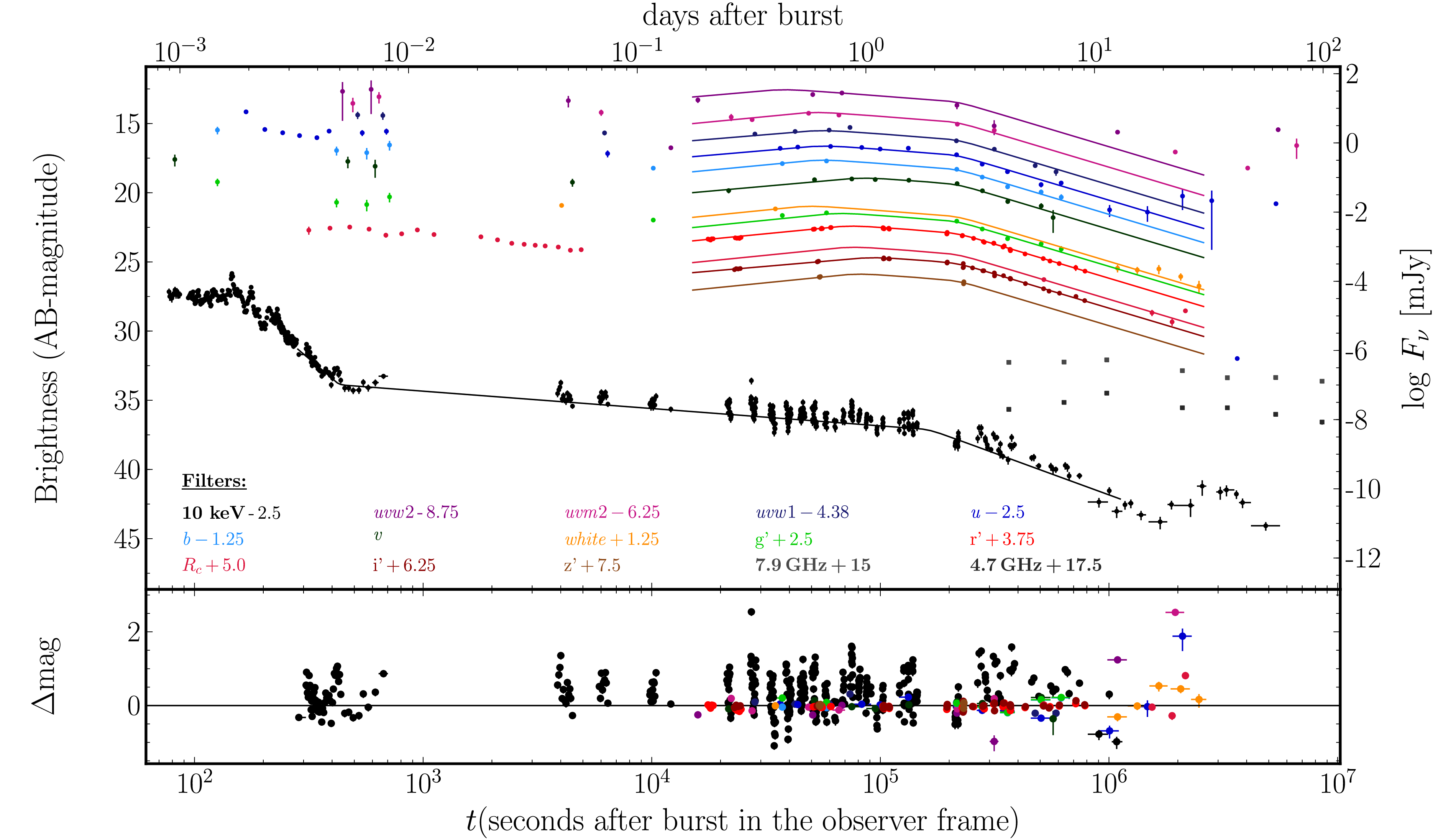}
 \caption{Combined X-ray, UV/optical/IR/radio lightcurves of GRB100814A.}
 \label{f1}
\end{center}
\end{figure*}

\begin{figure*}
\begin{center}
 \includegraphics[angle=-90,scale=0.42]{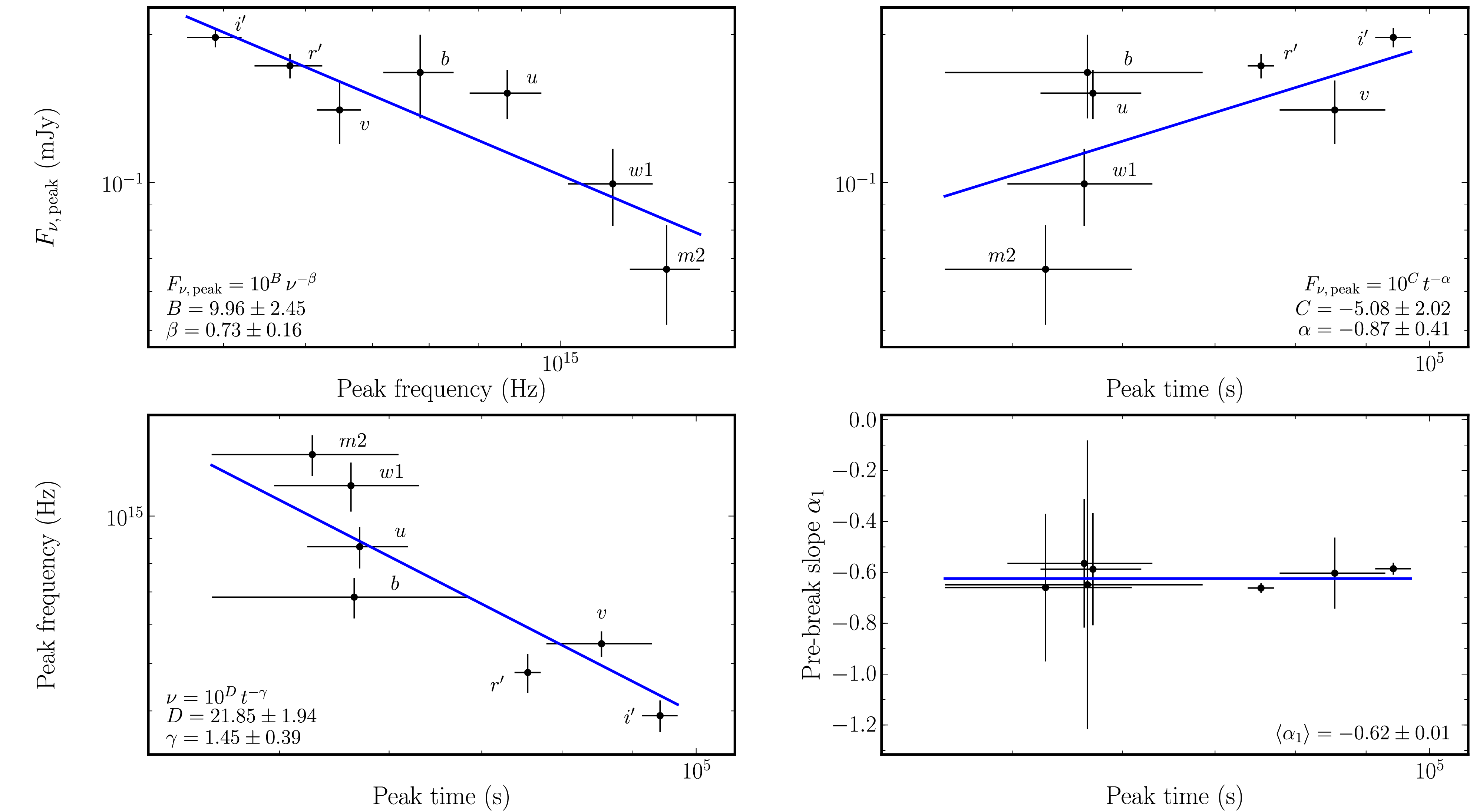}
 \caption{Correlations among peak time, peak flux density, peak frequency at the optical rebrightening.}
 \label{f2}
\end{center}
\end{figure*}

\section{Conclusions}

GRB10814A is an event with differing X-ray and optical behaviour, in particular displaying a late optical rebrightening which may be chromatic. This event joins a sample of bursts showing the same intriguing properties. We have tested three possibilities to explain this behaviour: presence of two jets, combination of reverse and forward shock, and an internal emission mechanism. We find that the first two scenarios require conditions which are unlikely or can account only for a few of the observed features, and the third scenario may give an explanation only by not making clear predictions.

\end{document}